# A multidomain relational framework to guide institutional AI research and adoption

VINCENT J. STRAUB, Alan Turing Institute, UK

DEBORAH MORGAN, Alan Turing Institute, UK and University of Bath, UK

YOUMNA HASHEM, Alan Turing Institute, UK

JOHN FRANCIS, Alan Turing Institute, UK

SABA ESNAASHARI, Alan Turing Institute, UK

JONATHAN BRIGHT, Alan Turing Institute, UK

Calls for new metrics, technical standards and governance mechanisms to guide the adoption of Artificial Intelligence (AI) in institutions and public administration are now commonplace. Yet, most research and policy efforts aimed at understanding the implications of adopting AI tend to prioritize only a handful of ideas; they do not fully connect all the different perspectives and topics that are potentially relevant. In this position paper, we contend that this omission stems, in part, from what we call the 'relational problem' in socio-technical discourse: fundamental ontological issues have not yet been settled—including semantic ambiguity, a lack of clear relations between concepts and differing standard terminologies. This contributes to the persistence of disparate modes of reasoning to assess institutional AI systems, and the prevalence of conceptual isolation in the fields that study them including ML, human factors, social science and policy. After developing this critique, we offer a way forward by proposing a simple policy and research design tool in the form of a conceptual framework to organize terms across fields—consisting of three horizontal domains for grouping relevant concepts and related methods: Operational, Epistemic, and Normative. We first situate this framework against the backdrop of recent socio-technical discourse at two premier academic venues, AIES and FAccT, before illustrating how developing suitable metrics, standards, and mechanisms can be aided by operationalizing relevant concepts in each of these domains. Finally, we outline outstanding questions for developing this relational approach to institutional AI research and adoption.

## 1 INTRODUCTION

Public institutions, such as government ministries and executive agencies, are increasingly making use of artificial intelligence (AI), particularly machine learning-driven (ML) systems, with the aim of improving service delivery and informing policymaking [19]. The advanced capabilities of these tools have prompted the recognition that we need new metrics, technical standards and governance mechanisms to evaluate and guide their use. However, while research on institutional AI, research related to the technical as well as ethical, social, political and legal implications of algorithms and computing in public administration, is now commonplace, most work arguably still fails to account for the diverse potential advantages and consequences of adopting AI in a public sector context. Instead, reflecting trends in socio-technical discourse more broadly, many contributions at premier conferences arguably tend to foreground only

Authors' addresses: Vincent J. Straub, vstraub@turing.ac.uk, Alan Turing Institute, British Library, 96 Euston Rd, London, UK, NW1 2DB; Deborah Morgan, Alan Turing Institute, British Library, 96 Euston Rd, London, UK, NW1 2DB and University of Bath, Claverton Down, Bath, UK, BA2 7AY; Youmna Hashem, Alan Turing Institute, British Library, 96 Euston Rd, London, UK, NW1 2DB; John Francis, Alan Turing Institute, British Library, 96 Euston Rd, London, UK, NW1 2DB; Saba Esnaashari, Alan Turing Institute, British Library, 96 Euston Rd, London, UK, NW1 2DB; Jonathan Bright, jbright@turing.ac.uk, Alan Turing Institute, British Library, 96 Euston Rd, London, UK, NW1 2DB.

a handful of topics, perspectives, concepts and methods [2, 5, 9, 14, 34], such as mathematical formulations of outcome fairness in ML applications [17], at the expense of others. How then do we ensure that future research on new metrics, technical standards and governance mechanisms better accounts for all the topics, concepts and methods potentially relevant to the institutional adoption of AI?

In this position paper, we focus on one theoretical issue, which we call the 'relational problem', that has arguably hindered scholarly efforts at the two premier conference venues for socio-technical issues, AIES and FAccT, to comprehensively study AI systems in an institutional context: fundamental ontological issues within the field have not yet been settled—including semantic ambiguity and, more significantly, a lack of clear relations between different topics, perspectives, concepts and methods (henceforth also abbreviated to 'terms and approaches'), leading to differing standard terminologies across subcommunities. We contend that this failure exasperates the prevalence of disparate modes of reasoning to assess institutional AI systems—such as "formalist algorithmic thinking" in computer science [14]—and contributes to the prevalence of conceptual isolation in the fields that study them including ML, human factors, social science and policy. After developing our argument, we propose a simple research and policy design tool in the form of a conceptual framework to organize terms and approaches across disciplines—consisting of three horizontal domains for grouping relevant concepts and related methods: Operational, Epistemic, and Normative.

Within the context of AIES and FAccT, the utility of our framework derives from the fact that it seeks to be discipline-agnostic; it aims to be instructive for individual policymakers and researchers studying institutional AI systems from a range of disciplines, both in helping with organizing concepts and methods and, more importantly, by drawing attention to whether all potential topics and concepts—by virtue of being relevant to one or more of the three proposed domains—have been accounted for. Our framework therefore aims to achieve two key aims: (1) disciplinary reach, i.e., bridge different subcommunities at AIES, FAccT and elswhere (ML, human factors, social science, policy etc.), and (2) provide impetus for an intellectual shift that reframes how researchers and key stakeholders (decision-makers, policy creators, advocates, etc.) think about and ultimately regulate institutional AI applications.

The rest of the paper is structured as follows. In Section 2, we motivate our argument and situate it against the backdrop of recent socio-technical discourse at AIES and FAccT. We then introduce our framework in Section 3 and illustrate how developing suitable metrics, standards and mechanisms can be aided by identifying and operationalizing relevant concepts across each of the proposed domains. Finally, in Section 4 we conclude by outlining key questions needed to develop a research agenda for advancing a relational approach to institutional AI research and adoption.

## 2  THE STATE OF SOCIO-TECHNICAL DISCOURSE

In this section we consider why certain topics, concepts and methods have been disproportionately studied in socio-technical discourse studied and consider the role played by unresolved ontological issues. Importantly, throughout this paper, the word 'method' is taken to mean the different technical or policy measures that may be used to evaluate and guide the use of AI systems (i.e., metrics, mechanisms, standards etc.), either by operationalizing a specific concept (e.g., classification accuracy in the case of performance) or combining a number of concepts into a qualitative framework (e.g., algorithmic impact assessment). Concepts are meanwhile understood both as an abstract idea that offer a point of view for understanding some aspect of experience (e.g., bias), and, relatedly, a mental image that can be operationalized (e.g., measurement bias). As such, a loose parallel can be drawn between our use of the terms concepts and methods, and the terms 'principles and practices' in AI Ethics discourse [22].

## 2.1 The Double-Edged Sword of AIES and FAccT

As more institutions move to employ AI systems in high stakes decision-making contexts like criminal sentencing, heightened attention has been drawn to the detrimental effects this can have—especially for marginalized and traditionally under-served groups—ranging from simple inefficiencies to major injustices [32]. Biased performance, inscrutable design and the uncritical implementation of complex AI applications have subsequently been identified, among others, as the main causes of these undesirable consequences [6, 10, 29]. More recently, the structural, historical and power disparities that permeate society and necessarily affect the design or adoption of technical systems have also received more attention [4, 5, 40]. Over the last five years, this discourse has matured, and a number of academic venues have become established platforms for computer scientists and scholars from other disciplines to raise awareness of these topics. Yet, this development has arguably been a double-edged sword. On the one side, it has put a bright and much-needed spotlight on socio-technical issues in AI. On the other, it has resulted in certain topics and methods receiving considerable attention, at the expense of other ideas and challenges. This unequal emphasis on particular topics has also characterized the growth of the two premier conferences, AIES and FAccT, which we focus on herein.

Officially, AIES and FAccT seek to consider the ethical ramifications of AI systems (including those used in public administration and social service provision) and their impact on human societies, address general questions that consider perverse implications, distribution of power, and redistribution of welfare and ground research in existing legal requirements. In practice, however, there has been a disproportionate focus on a handful of narrow topics and methods. [17] diagnose this troubling trend most clearly in their four-year analysis of FAccT, finding that there has been an out-sized focus on, among other topics, "quantitative work on fairness, displacing discussions about broader AI policy and governance, both within and across years". This is in spite of the more general way that FAccT defines its aims. Similarly, [5] find that while the goals of the majority of contributions to AIES are often commendable, "their consideration of the negative impacts of AI on traditionally marginalized groups remained shallow", leading them to conclude that there is an overall inadequacy of scholarship in engaging with perspectives that are "not a part of the standard".

To be clear, it is to be expected that different subcommunities and disciplines study and thereby value different topics, concepts and issues. This is, after all, the point of specialization. A problem occurs, however, when fields are meant to be united in studying the same topic (i.e, socio-technical issues in AI systems) but do not acknowledge these differences and fail to integrate ideas from their peers. The result is the use of divergent terminologies and the exasperation of knowledge silos, meaning terms like 'fairness' and 'discrimination' are understood differently by ML researchers than by HCI or AI Ethics scholars [40]. In some cases, this even means that certain highly valued topics and approaches may become embedded in supposedly value-neutral and universally beneficial research, e.g., generalization, quantitative evidence, and efficiency in the case of ML [4]. This situation threatens to create great challenges for effective understanding, dialogue, and integration between disciplines and academic subcommunities.

## 2.2 Ontological Issues and The Relational Problem

We are by no means the first to consider whether certain terms and approaches have dominated socio-technical discourse at AIES and FAccT. In fact, there have been a growing number of calls from within the scholarly community to diversify the number of topics studied within recent years. In prior work examining trends in recent proceedings, scholars have offered in-depth analyses of why and how certain topics, such as fairness [2, 7, 15, 23, 30], have been researched more than others [3, 5, 17, 40]. For instance, [40] argue that, because of corporate capture, i.e., conflicts of interest, conference contributions "frequently limit their gaze to the 'technical' part of 'sociotechnical'—the level of data, metadata, or models" and reduce complex concepts like fairness to "dimensions of arbitrary narrowing that both obscure and reproduce structures of social injustice". Similarly,[35] contend that, by abstracting away the social context in which these systems will be deployed, "fair-ML researchers miss the broader context, including information necessary to create fairer outcomes, or even to understand fairness as a concept". [5] meanwhile offer a more sociological explanation, arguing that existing scholarship has been influenced by the centralization of power between highly cited researchers, tech companies, and elite universities, resulting in topics and concepts related to oppressive social structures, the distribution of power, and harm receiving less attention.

Yet, while it is important to acknowledge that sociological factors and economic incentive structures (i.e., scientific funding) clearly influence *what* different topics researchers study [39], here we wish to focus on an altogether more theoretical factor relevant for understanding differences in *how* researchers study the same topic. Specifically, our main contention is that discrepancies and inequalities in the terms and concepts that researchers employ to study institutional AI are, in part, connected to the fact that fundamental ontological issues within socio-technical discourse have not yet been settled. That is, there are semantic ambiguity problems, specifically, a lack of agreed upon definitions for key terms, a lack of clear and consistent relations between topics, concepts and related methods, and differing standard terminologies across subcommunities. [17] draw attention to this problem in passing, noting, for instance, that universally agreed upon definitions are even lacking for foundational terms like 'AI'.

Although the lack of commonly agreed upon definitions within fields is an ongoing problem, we contend that the most troubling development is the failure of research efforts to establish clear relations and make connections between different topics and concepts—especially between those that are easily quantifiable and those that are not. One recent survey of ML research, for instance, found that terms related to user rights and ethical principles, like interpretability, privacy and non-maleficence, "appeared very rarely if at all" compared to performance or efficiency, and "none of the papers mentioned autonomy, justice or respect for persons" [4]. This is despite the fact that these represent topics that are clearly also important when considering the application of ML, as scholars in AI Ethics have long shown. In the case of technical work on measures and methods to improve fairness in AI systems [2], for instance, [23] contend that most so-called 'fairML' efforts happen in isolation and lack "serious engagement with philosophical, political, legal and economic theories of equality". Instead, researchers oversimplify or 'level down' the broader topic of distributive justice into a single evaluation metric that attempts to operationalize fairness, while other topics and approaches (e.g., algorithmic impact assessments that also consider priority and welfare) are minimally discussed or ignored altogether. This results in semantic ambiguity, as concepts like fairness can have multiple definitions and can mean very different things depending who you ask [25].

It is important to re-emphasize that certain concepts and methods have necessarily received disproportionate attention given they currently exert an out-sized influence on scientific progress, public institutions or society more broadly. To stay with the obvious example, ML is actively being used or trialed in myriad different applications and public administration contexts [11], including in healthcare, policing, criminal justice and other so-called high stakes domains [33]—where fairness is inherently important. As such, it is to be expected that ML, fairness and consideration of these high stakes domains has been a considerable focus of proceedings at AIES, FAccT and broader societal discourse at large. The issue we wish to stress is that most work on these topics does not effectively relate the terms and approaches used to other, perhaps less well-studied but potentially equally relevant and important terms and approaches. This results in rich-get-richer and echo-chamber system dynamics in institutional AI research as whole, whereby certain perspectives of important topics (i.g., mathematical definitions of outcome fairness) dominate discussion. Alongside fairML and other related topics (e.g., Explainable AI [20]), some scholars have started to draw attention to this issue in the context of specific subfields or disciplines. [14], for instance, argue that computer scientists, as a result of adopting a formalist mode of reasoning, often do not fully engage with other disciplines when considering the social and political contexts of AI systems. Drawing on studies of sociotechnical systems in Science and Technology Studies, [35] similarly argue that when researchers treat fairness and justice as terms that have meaningful application to technology separate from a social context, they make a category error, or as they posit, an 'abstraction error'. This is because fairness is a property of social and legal systems like employment justice, not a property of the technical tools within a system.

Overall, this ontological failure to explicitly connect terms and approaches, which may be called the 'relational problem'—as it mirrors debates in AI Ethics on how to reframe concepts in relation to those affected [3]—will, if left unaddressed, arguably only worsen the existence of conceptual isolation in the fields that study institutional AI adoption including ML, human factors, social science and policy. In the context of AIES and FAccT, this will in turn likely make it easy for certain concepts (e.g., fairness) and methodological formulations (e.g., mathematical) favored by popular subfields (e.g., ML) to continue dominating discussion. As a result, other topics and perspectives may be pushed further to the margins of discourse [5] and any definitional consensus may be better described as manufactured rather than genuine [40]. Most importantly, it ultimately means that the development of new metrics, technical standards and governance mechanisms to guide the adoption of AI in public administration ends up reflecting only a small subset of perspectives and concepts pertinent to the complex reality of institutional AI. As a consequence, we are left with, at best, a distorted view of the implications of adopting AI systems and, at worst, the neglect and perpetuation of real-world algorithmic harms and injustice that affect historically disadvantaged or marginalized groups the hardest.

## 3 A RELATIONAL MULTIDOMAIN FRAMEWORK FOR INSTITUTIONAL AI

How then do we address these ontological issues and ensure that future socio-technical research on new metrics, technical standards and governance mechanisms better reflects all the terms and approaches potentially relevant to the institutional adoption of AI? What is arguably needed is a change in how researchers and policymakers conceptualize the application of AI in institutional contexts to begin with. That is, to move beyond current disparate modes of reasoning, which each do not fully account for the realities of algorithmic impacts, requires a fundamental shift—from a single lens to multiple perspectives—in how to think about all relevant topics, concepts and methods—be it outcome fairness, welfare, performance or accuracy metrics—and how to link them to each other. Given that many concepts employed to discuss socio-technical issues at conferences like AIES and FAccT are fundamentally multi-faceted or discipline-specific, we do not wish to propose new definitions or prescribe which specific terminologies or modes of reasoning should be used. Rather, we use the rest of the paper to propose, as a starting point, a simple policy and research design tool in the form of a conceptual framework to organize terms and approaches across fields.

Our framework consists of three discipline-agnostic domains for grouping relevant concepts and related methods that each have a distinct thematic and semantic scope. We label these: Operational, Epistemic, and Normative. The main aim of our framework is to achieve two specific aims: (1) disciplinary reach, i.e., bridge different perspectives (CS, human factors, social science etc.), and (2) provide impetus for an intellectual shift that encourages researchers and key stakeholders (decision-makers, policy creators, advocates, etc.) to think about institutional AI systems more holistically. Our overarching goal is to offer a way to organize disparate socio-technical research outputs into general thematic categories, making it easier to align and integrate efforts from different scholarly subcommunities. Below we first introduce and define the domains before discussing how they can be integrated and unified into a single framework.

### 3.1 Grouping Socio-Technical Topics into Three Domains

Our framework is ontological in the sense that is it composed of three domains or meta-concepts that aim to act, both individually and collectively, as guides for researchers to relate and connect different terms and approaches within socio-technical discourse to each other. Importantly, the concepts and terms that can be grouped into one of the three domains are not synonymous but we assume that they are all used to discuss institutional AI systems in a similar way. As such, each domain can loosely be said to function as a semantic field, a set of words related in meaning (i.e., terms used to study institutional AI systems), and are defined by their unique thematic focus, or what we label as 'scope'.

To theorize and define the scope of the three domains we took inspiration from three strands of work. Firstly, social science and human factors research that emphasizes the behavior and beliefs of human agents in influencing the performance of technical systems including AI applications (e.g., [31]). Secondly, computer science work on reasoning about the knowledge-related properties of a technological system [8]. And thirdly, the insight of moral philosophers that questions about the social implications of socio-technical systems including AI applications depends on political decisions about normative issues [12].

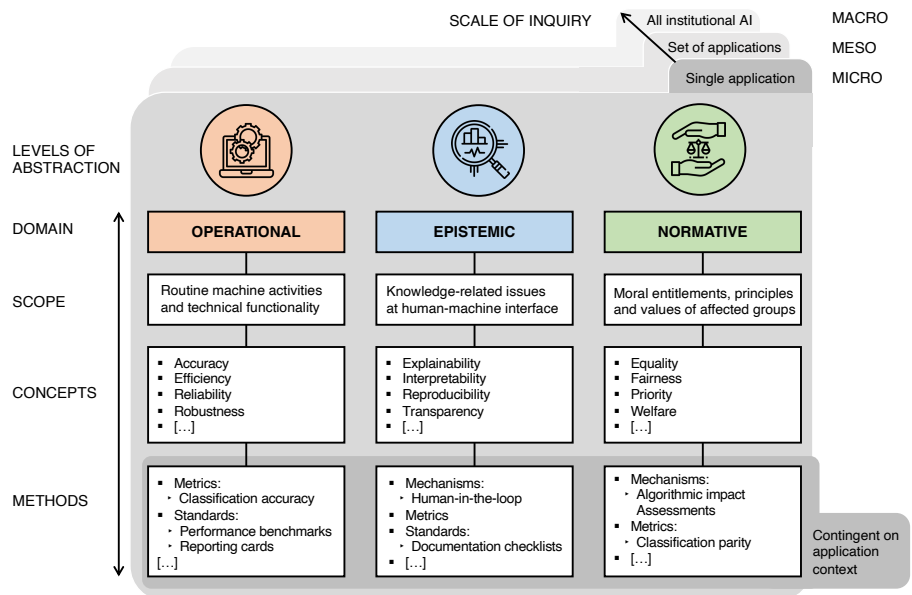

Fig. 1. Graphic representation of our framework for grouping relevant concepts and related methods relevant to institutional AI adoption. The framework consists of three domains that each have a distinct thematic and semantic scope: Operational, Epistemic, and Normative. These domains can be used to consider AI at all three scales of analytical inquiry: a single AI application (micro), sets of similar applications (meso) and all institutional AI applications considered as a single class (macro). Listed concepts and methods are for illustrative purposes and not exhaustive; example methods (e.g., the metric of classification accuracy) are taken to be contingent on the local application context.

*3.1.1 Operational Domain.* The operational domain aims to represent the terms and approaches related to the routine activities and functionality of institutional AI systems [29]. Its scope is meant to capture concepts that are mainly but not exclusively defined, operationalized and studied in a technical, applied context. More specifically, it is meant to enable researchers to categorize into a single category all relevant concepts that can be employed both as an abstract idea (e.g., 'accuracy') and easily operationalized to quantitatively measure a specific performance attribute of a particular institutional AI system (i.e., 'percentage of correct predictions'). As such, the common characteristic of all the concepts and related methods in this domain, regardless of how they are operationalized, is an emphasis on describing specific functional requirements or attributes of institutional AI systems. This reflects what is arguably unique about the operational domain, namely, it aims to draw attention to concepts that can be conveniently specified in a common technical (mathematical) language or easily quantified, allowing for the succinct description and comparison of specific models or applications.

*3.1.2 Epistemic Domain.* The scope of the epistemic domain aims to capture knowledge-related terms and approaches connected to a particular AI system or institutional AI in general. That is, the epistemic domain is meant to help researchers and policy-makers group together concepts that seek to describe properties which pertain to the interface between AI applications and human actors. Both in terms of the knowledge, beliefs, and intentions of those using AI applications (e.g., a desire for transparency), and the internal properties of the system itself (e.g., its interpretability). Given that AI systems represent a step-change from earlier ICTs due to their increased technical complexity, among other factors, epistemic domains concepts are likely also useful in delineating different types of AI systems. Relatedly, while the domain seeks to capture terms that are often employed analytically to highlight epistemic issues about institutional AI systems (a lack of reproducibility, openness etc.), it is also meant to account for and help organize methods and concepts operationalized in a technical manner to improve human knowledge of a system (e.g., explainability, interpretability).

*3.1.3 Normative Domain.* The meaning and uses of concepts in the normative domain, the final domain we propose, collectively relate to the entitlements, values and principles of political morality that stakeholders and affected groups hold towards a particular AI application or institutional AI in general. The term 'political morality' is used here to refer to normative principles and ideals regulating and structuring the political domain. In the context of institutions, stakeholders may be said to include system developers (e.g., designers, engineers, and domain experts), those who manage and operate them within the public sector (e.g., decision-makers, policy creators, advocates), and end-users affected by the AI system (i.e., individual citizens or specific groups). In some cases, this may include large parts of society [28], as AI systems can increasingly result in individual and collective harms [38]. We anticipate that the normative domain primarily covers concepts that can be understood in two ways. Firstly, those used in a practical ethics sense, such as in bioethics, to stress the values that underlie the safeguarding of individuals (e.g., 'non-maleficence'). Secondly, those used in a legal framing, as in human rights discourse, to discuss the set of entitlements due to all human beings under the rule of law for a particular jurisdiction [18] (e.g., justice). Importantly, while this means concepts and topics in the normative domain may appear to only be relevant to particular disciplines (i.e. AI Ethics), this is by no means the case as normative domain concepts also need to be operationalized by computer science researchers if we are to move from principles to practices [22].

## 3.2 Integrating the Dimensions

The operational, epistemic and normative domain are each meant to act as independent analytical categories that can be used to help researchers focus on a particular aspect of institutional AI (e.g., knowledge-related issues) and make connections between related concepts (e.g., explainability, transparency, etc.). However, the real utility of this relational approach comes to the fore when each of the three domains are integrated and unified into a unified framework. Figure 1 provides a graphic representation of this. When each domain is considered together in this polycentric manner, the emphasis is on the conceptual need to always think horizontally when proposing new methods or discussing socio-technical issues in AI systems including institutional AI applications. That is, rather than expecting scholars to employ every single concept in a particular domain when discussing how to evaluate an AI system, the framework aims to encourage researchers to connect concepts and methods across the three domains. In practice, this means that ML researchers currently working on fairML, for instance, are reminded to consider how fairness relates to other normative domain concepts like welfare and take into account the need to consider epistemic domain concepts like interpretability and reproducibility, alongside accuracy, efficiency and other operational domain concepts. Similarly, the framework reminds AI Ethics and policy scholars working on transparency to link their work laterally to other epistemic domain concepts and not ignore operational domain concepts like robustness and reliability.

Crucially, the framework does not seek to prescribe which specific concepts in each domain are most important or which related methods are most useful and relevant. As such, each of the example concepts and methods discussed herein and listed in Figure 1 are cited primarily for illustrative purposes. Different disciplines and academic subcommunities will, as already discussed, rightly study and value particular topics, and certain terms and approaches necessarily receive more scholarly attention. In light of this, the framework aims to act primarily as a discipline-agnostic design tool that can help researchers and policymakers organize diverse concepts across fields and motivate them to adopt a more relational, holisitic approach. Hence, we do not attempt to list all of the concepts and methods that fall into each domain. Rather, by outlining and specifying the scope of each domain, we encourage researchers that study institutional AI including scholars in ML, human factors, social science, policy and AI ethics, to start identifying and connecting additional concepts themselves. While we have tried to define the scope of each of the domains in a narrow enough way so that any socio-technical concept falls into a single domain, certain complex concepts may naturally span more than one domain. 'Accountability', for example, is often defined too imprecisely and can pertain to a variety of values, practices, and measures. It is considered by some scholars as a necessary feature of a trustworthy AI system, while others argue that only humans can be accountable [26]; depending on whether they are defined in function-based terms or not, concepts like accountability or trustworthiness [37] may thus be considered to be an operational and or normative domain concept.

A further important clarification pertains to how various methods may relate to particular concepts. More specifically, while foundational concepts like accuracy are more or less taken to have universal relevance when it comes to the institutional adoption of AI systems, differences in moral values [1], including with regards to the AI use case [27], and practical contextual factors (i.e., the type and number of systems that are institutionally adopted) additionally means that certain concepts may in practice be more important than others. As such, the exact concepts and related methods which are most relevant to each domain are necessarily contingent on the application context. That is, while we envisage that our framework can be used to study and evaluate AI systems at different scales of analytical inquiry, when the object of study is a specific AI system designed for a local institutional application context, we nevertheless anticipate that the most relevant methods to operationalize particular concepts may change. For instance, data documentation checklists and model reporting cards [21] may be considered sufficient when seeking to apply operational and epistemic domain concepts like reliability and interpretability, respectively, to understand the adoption of a recommender system to provide suggested links on a local government domain. However, if a similar system is used by a national healthcare provider to recommend medication, additional methods may be necessary (e.g. mechanisms like human-in-the-loop operating protocols).

### 3.3 Applying the framework in practice

Overall, our framework is intended to help researchers and policymakers within various fields engaged in studying and regulating institutional AI systems, such as AI-assisted decision support systems or criminal justice tools. Specifically, it is meant to act as a starting point for conceptualizing the desired attributes of AI systems, and thus purposely aims to foreground the need to integrate ideas, alongside being applicable to various real-world examples of AI systems, and remaining stable and useful over time as a conceptual model [24]. In context of AIES and FAccT, the framework's utility therefore derives from the fact that it seeks to be discipline-agnostic; it aims to be instructive for individual researchers studying institutional AI systems from a range of disciplines, both in helping with organizing terms and approaches, and, perhaps more importantly, by drawing attention to whether all potential intellectual and moral perspectives—by virtue of being relevant to one or more of the three proposed domains—have been accounted for.

Despite the theoretical nature of our framework, we anticipate that it can practically help address some of the ontological issues we outline, such as as the need to bridge quantifiable and non-quantifiable terms and concepts, when it is viewed as a simple policy or research design tool. That is, we contend that the framework can be used as a strategy to help researchers go about deciding which terms and approaches are relevant for studying a single or set of AI systems and ensuring they assess these from multiple perspectives. This can be achieved by relying on the four levels of abstraction (see Figure 1) to deductively guide the process of conceptualization. In other words, after first relying on the three meta-concepts (domain) to ensure all types of concepts covering different thematic areas (scope) are accounted for, researchers can then choose particular terms (concepts) that are most appropriate to the system under consideration, before finally operationalizing these (methods), depending on the application context.

As an example, consider the use of a recommendation system, special-purpose software designed to suggest content to a user of an online service, in an institutional context, such as for suggesting links to citizens on a pubic domain government website. Although recommendation systems like Google Search's autocomplete function and Amazon's recommendations for related products are well-known examples of AI systems, they carry a number of ethical implications and the use of similar systems within public institutions adds another layer of ethical complexity [16], as is the case for the UK's GOV.UK, which uses machine learning to guide guide users through complex service journeys [36]. To understand and regulate such public service recommender systems, the framework encourages authors to consider operational as well as epistemic and normative topics, ensuring they are situated within the fairness, accountability, and transparency discourse [13]. Specifically, it reminds authors to consider how epistemic topics like explainability, interpretability and reproducibility may be important for ensuring a system is democratic. Similarly, it reminds authors to also consider how fairness, equality, and welfare may need to be considered to ensure the system meets legal accessibility requirements (e.g., can be accessed on legacy devices) and does not infringe privacy concerns by relying on user data or provide biased outputs. Measuring and evaluating each of these criteria may in turn involve multiple methods (i.e., metrics, standards, and mechanisms) that will be contingent on the application context. For instance, for operationalizing normative concepts like equality, justice and fairness, researchers and policymakers will need to rely on the fundamental rights enshrined in law for the particular jurisdiction where a system is being implemented; in some cases these may be more or less universal (e.g., the prohibition of discrimination).

## 4 MAPPING A RESEARCH AGENDA FOR A MULTIDOMAIN APPROACH TO INSTITUTIONAL AI

Our conceptual study has primarily aimed to shed light on theoretical, specifically, ontological issues in socio-technical discourse, focusing in particular on contributions to AIES and FAccT, and considered how we might begin to resolve them. While these conferences continue to be dominated by a subset of topics and methods, there are signs of a shift, evidenced, for instance, by the gradual but significant increase in legal, social science, and ethics papers over the years, alongside ML papers about fairness [17]. Yet, the fundamental relational problem we described in Section 2 will arguably remain until scholars start actively integrating more perspectives, concepts and methods from their peers.

While we hope our framework can enable all researchers at AIES, FAccT and elsewhere to adopt a more holisitic approach to conceptualizing and evaluating institutional AI systems, we wish to stress that it is only a first step. We must not only consider the use of multiple domains to assess and evaluate institutional AI systems but also understand how each works together. As such, we have identified 10 key outstanding questions that we anticipate will be key for developing this multidomain relational approach to institutional AI research and adoption:

(1) Do we need to further delineate and operationalize the operational, epistemic, and normative domains as tangible concepts, or is it enough for these to act as abstract categories of analysis?
(2) To what extent does there need to be scholarly consensus on how we decide whether concepts fall into a particular domain and not into a different one?
(3) Should the importance of different concepts and metrics in a particular domain be considered? And if so, how?
(4) How much attention and focus on one domain at the expense of the other domains is acceptable?
(5) Is it of value to consider how we can move to unite each of the domains into a single category?
(6) How can lessons across domains be captured to develop their definitions?
(7) Which methods in each domain are least contingent on the application context?
(8) How can we decide which methods are most appropriate for operationalizing a particular concept?
(9) What other unique domains may exist that capture enough additional concepts to be worthy of inclusion as new domains?
(10) How can be empirically quantify the strength of relations between different concepts and methods?

## 5 CONCLUSION

This position paper has considered why most research and policy efforts aimed at understanding the implications of institutional AI tend to prioritize only a handful of ideas, and how this relates to the state of socio-technical discourse more broadly. Specifically, we have sought to highlight one fundamental theoretical issue, which we call the relational problem, that has arguably hindered scholarly efforts at two premier socio-technical conference venues, AIES and FAccT, to comprehensively study AI systems: fundamental ontological issues within the field have not yet been settled— including semantic ambiguity and, more significantly, a lack of clear relations between different topics, perspectives, concepts and methods, leading to differing standard terminologies across subcommunities. We contend that this failure has contributed to the prevalence of conceptual isolation in the fields that study them including ML, human factors, social science and policy, among others. In response, we have offered a way forward by proposing a simple policy and research design tool in the form of a conceptual framework to organize terms across fields—consisting of three horizontal domains for grouping relevant concepts and related methods: Operational, Epistemic, and Normative.

The main contribution of our research is providing a first step for those studying institutional AI to connect topics and consider whether all relevant topics and concepts have been accounted for. Future work will benefit from further considering the ontological ontological and epistemological underpinnings of the relational problem. While we have focused on understanding how the existence of ontological issues in socio-technical discourse ensures research remains fragmented, several factors may explain how this arises to begin with, relating to the relative newness of the field, the transdisciplinary nature of the work, the sociopolitical dynamics of academic research, the influence of industry, to name a few. A fruitful avenue of inquiry will be to consider each of these interact, what other plausible contributors are, and what the implications are for applying the framework we put forth.

In closing, we hope our contribution benefits the AIES and FAccT community by facilitating a constructive dialog around the challenges we face as a diverse, interdisciplinary field aiming to address sensitive, high-stakes socio-technical issues that will only grow in magnitude and significance in the years to come. In these hotly contested spaces with no clear answers, by analyzing these problems across three domains, we contend that we are able to more clearly see the many interacting parts at play, in order to create more functional, ethically sound institutional AI systems.